\documentclass[onecolumn,11pt]{article}
\def \degC {~^{\rm{o}}}

\usepackage{graphics}
\usepackage[square,comma,longnamesfirst,numbers,sort&compress]{natbib}
\usepackage[pdftex]{graphicx}
\usepackage[in]{fullpage}

\usepackage{amssymb,amsfonts,amsmath}
\usepackage[nomarkers]{endfloat}

\begin{document}
	
\title{High temperature viscosity measurement system and viscosity of a common dielectric liquid}

\author{Enis Tuncer\\
	{\textit{General Electric Global Research, Niskayuna 12309 New York USA}}}

\maketitle

	    
\begin{abstract} 
A device to measure viscosity of dielectric oils was developed. The device is an inset to an autoclave system where the temperature and the pressure could be controlled. It consists of a magnet and a coil. The magnet was externally driven to a location in the device and then let fall with removal of the driving force. The time-to-fall was recorded and converted to viscosity for a known dielectric fluid. The device is capable of measuring viscosities up to 400$\degC$C and 5000psi, which are the limits of our autoclave at the moment.

  \begin{itemize}
  \item[{\em Keywords}]{High temperature, high pressure, oil viscosity, liquid dielectrics, autoclave}
\item[{\em Disclaimer}] This report was prepared as an account of work sponsored by an agency of the United States Government. Neither the U.S. Government nor any agency thereof, nor any of their employees, makes any warranty, express or implied, or assumes any legal liability of responsibility for the accuracy, completeness, or usefulness of any information, apparatus, product, or process disclosed, or represents that its use would not infringe privately owned rights. Reference herein to any specific commercial product, or process, or serviceable by trade name, trademark, manufacture, or otherwise does not necessarily constitute or imply its endorsement, recommendation, or favoring by the U.S. Government or any agency thereof. The views and opinions of authors expressed herein do not necessarily state of reflect those of the U.S. Government or any agency thereof. 
  \end{itemize}

\end{abstract}
\section*{Introduction}
Viscosity data on liquid dielectrics are important in design of high temperature electrical equipment such as electrical submersible pumps for geothermal energy applications. There are numerous methods to measure viscosity of liquids as reviewed by \citet{Brooks2005}. The proposed method is based on a linear motion of a magnet in a field concept and gravitational force. It can measure the viscosity as a function of temperature and pressure. Using the method several liquid dielectrics are characterized for 330$\degC$C and 4400psi application.

\section*{Concept}
The main purpose of developing the presented system is to be able to measure the viscosity of a dielectric oil as a function of temperature (up to 400$\degC$C) and pressure (up to 5000psi). Safety in the experiment is one of the major concerns due to flash point of oils and the required pressure. However, a commercially available autoclave would provide the right experimental conditions, but one needs a viscosity meter for the characterization. We have designed a device  composed of a coil to generate the magnetic field and a magnet, which moves in up in the coil. The coil was built around a glass structure. Later the magnet was let fall in the liquid medium. The time-of-fall was measured between the current shut-off of the field and the impact of the magnet to the bottom of the glass tube. The measured time was correlated to the viscosity by using a transfer function between the time-of-fall and viscosity for a known oil at low temperatures ($T<120\degC$C). 

\begin{figure}[htp]
  \centering
  \includegraphics[width=\linewidth]{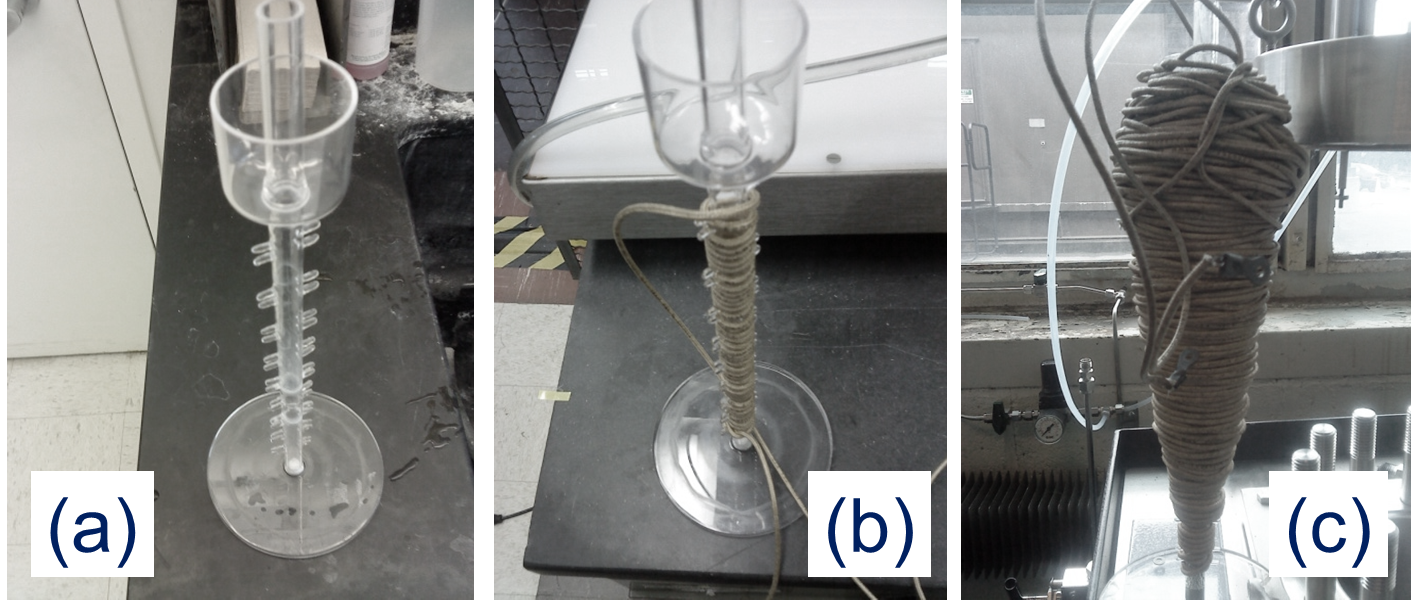}
  \caption{Images of the concept. (a) The glass tube with notches for wounding the high temperature 10 gage heater wire. and the magnet coil designed to drive a magnet}
  \label{fig:imag1}
\end{figure}

The concept pictures are shown in Fig.~\ref{fig:imag1}. A glass tube structure (0.5in diameter) about 13.25in high was designed and built due to high temperature requirements. Small rods were put in 1in spacings to help the winding process to form the magnetic coil. A container at the top part of the tube was used to capture any oil in case of its expansion with temperature. The picture from the start of the coil build-up and the final structure are shown in Fig.~\ref{fig:imag1}b and \ref{fig:imag1}c, respectively. The field coil was built in a cyclone-shape with increasing number of turns toward the top of the tube due to required driving force for the magnet. The height of the cyclone is about 12in. Once the magnet is inserted in the tube, and the coil is energized with a direct-current, the magnet levitates. (a Sorensen DCR 300-9B power supply with 400V/10A capability was used to apply the current to the coil). The coil resistance is 3.812$\Omega$ measured with a Agilent 34410A digital multimeter. The device was inserted inside an autoclave with electrical ports. The pressure and temperature inside the autoclave could be controlled in the chamber. In the current tests the charging pressure at room temperature was set to 400psi and the measurement were taken without changing the initial pressure in the system; observe that due to increase in temperature the pressure in the chamber increases slightly.

\begin{figure}[htp]
  \centering
  \includegraphics[width=\linewidth]{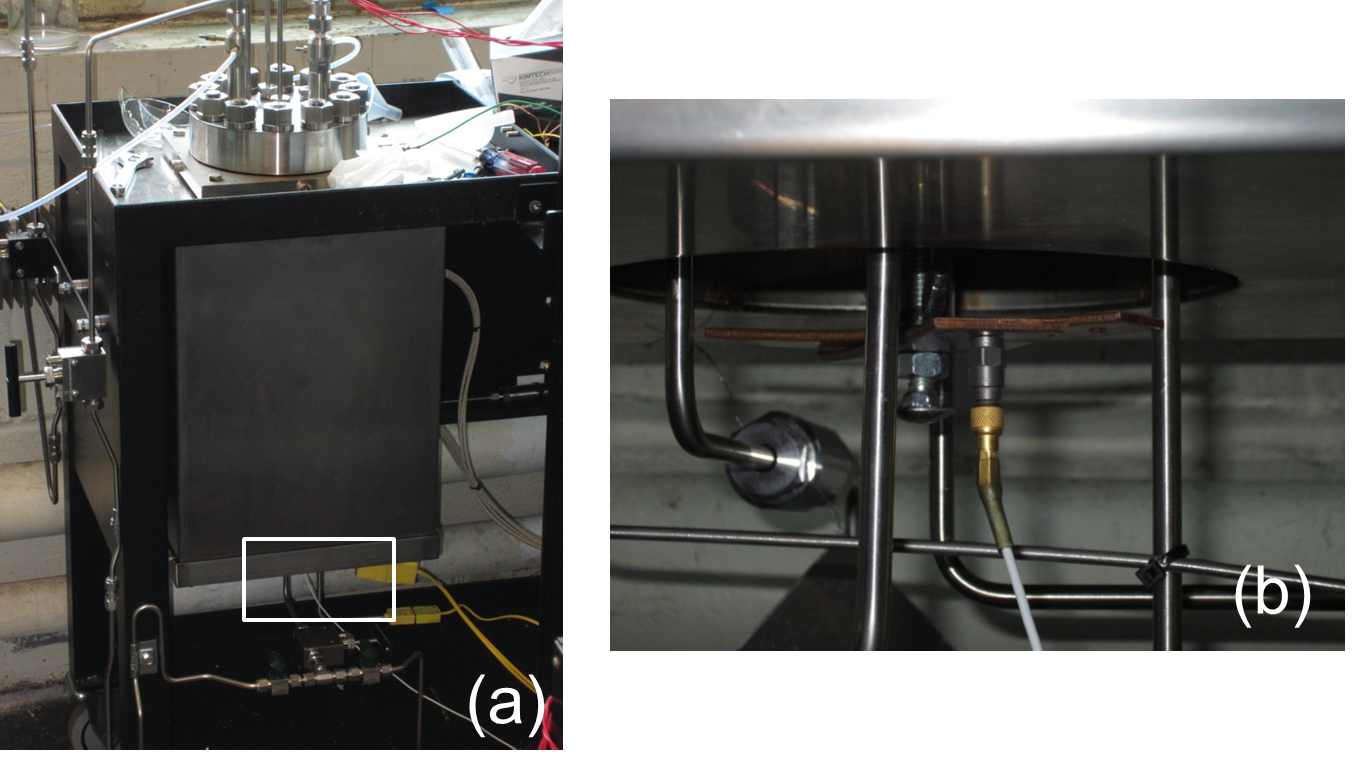}
  \caption{(a) Autoclave and (b) the position of the accelerometer. The accelerometer is located at the bottom of the autoclave. The rectangular region in (a) is enlarged in (b). The accelerometer is attached on a cantilever for improvement of the magnet impact.}
  \label{fig:imag2}
\end{figure}

\begin{figure}[htp]
  \centering
  \includegraphics[width=\linewidth]{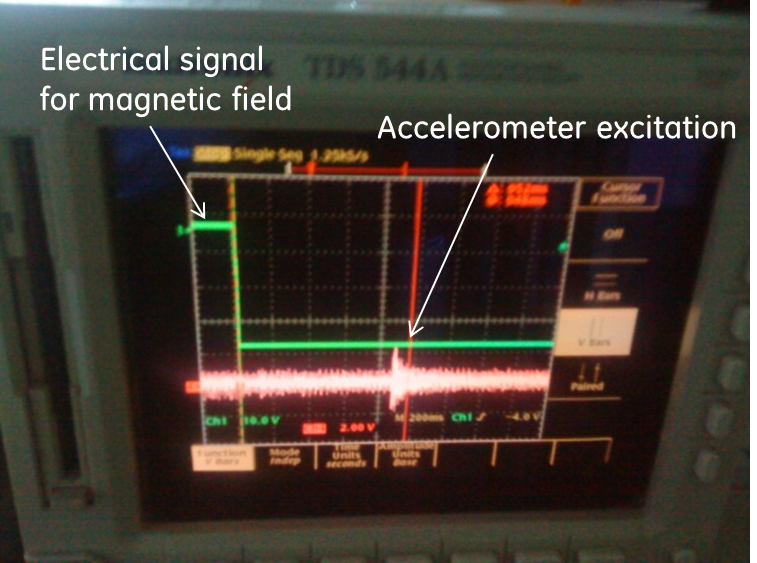}
  \caption{An example of the measurement on the scope screen. The turn-off current point and the signal from the accelerometer are shown. The time between the current turn-off point and the impact is registered as the time-to-fall for the magnet that is later used to estimate the viscosity.}
  \label{fig:scope}
\end{figure}

Magnet selection is important due to the De-magnetization properties of magnetic materials above their Curie temperature $T_c$; we need to operate below $T_c$. First a neodymium magnet was employed, however, its $T_c$ is about 310$\degC$C-400$\degC$C depending on the composition; this range is of the order of our highest temperature in the autoclave. Therefore, alnico magnet (composed primarily of aluminum, nickel and cobalt) is employed; $T_c$ about 800$\degC$C. The alnico magnet is 1/4in thick and 3.5in high (it is about 20g). Application of the current at room temperature causes the magnet to raise 7in with 10A applied to the coil. The duration of the magnet at the top position is kept less than 15s to avoid any additional heat generated by the coil.  After removal of the the applied current the magnet drops. The impact of the falling magnet creates a vibration in the autoclave that is registered using an accelerometer as shown in Fig.~\ref{fig:imag2}. The accelerometer is connected to the bottom of the autoclave through a  cantilever to improve its response; without the cantilever the signal-to-noise is not satisfactory. The signal from the accelerometer is increased using a home-made operational amplifier. The time-to-fall is measured with an oscilloscope as shown in Fig.~\ref{fig:scope}. The time of the current switch-off and the accelerometer trigger response are used for the time-to-fall $t$ value. Observe the decaying accelerometer signal in the scope. For air time-to-fall measurements the time is 200$\pm3$ms, which is less than 2\% accuracy. Measurements were performed every 2min while the temperature of the autoclave was increased slowly about 1$\degC$C/min.

\section*{Testing a conventional dielectric liquid}
Because of environmental restrictions new oils are introduced for electrical insulation (dielectric) applications. With increasing temperature requirements in transformational power equipment these new as well as old oils needs to be qualified for the specific application. For high temperatures ($T>250\degC$C), there are a few specific applications in the oil and gas industry that require coolants to keep the operating apparatus running efficiently without causing any problem in their insulation. The dielectric liquids therefore act both as an electrically insulating medium and also as a coolant for the system. 

\begin{figure}[htp]
  \centering
  \includegraphics[width=\linewidth]{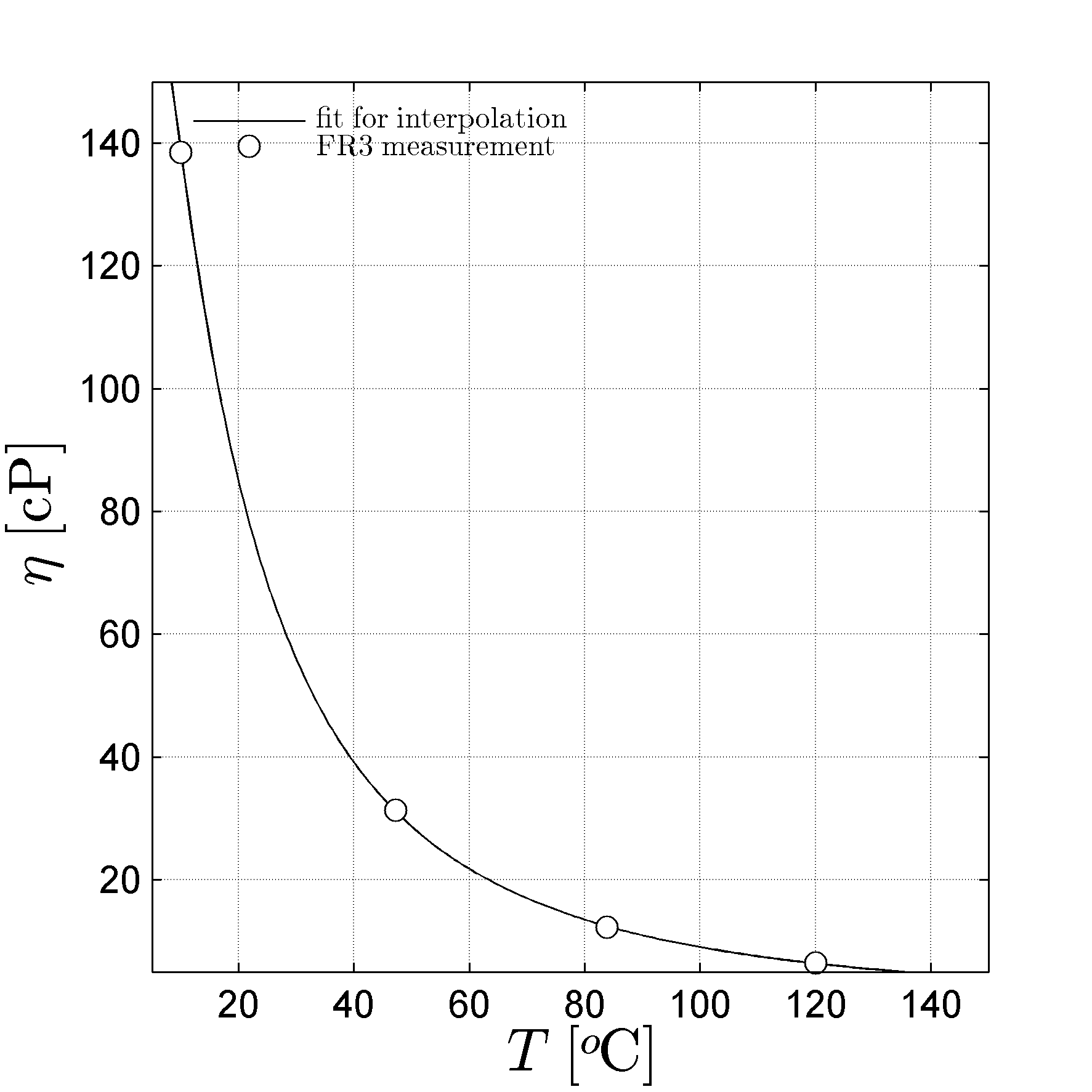}
  \caption{Viscosity of FR3\texttrademark as a function of temperature. The solid  line is used for the transfer function.}
  \label{fig:etaVStemp}
\end{figure}

\begin{figure}[htp]
  \centering
  \includegraphics[width=\linewidth]{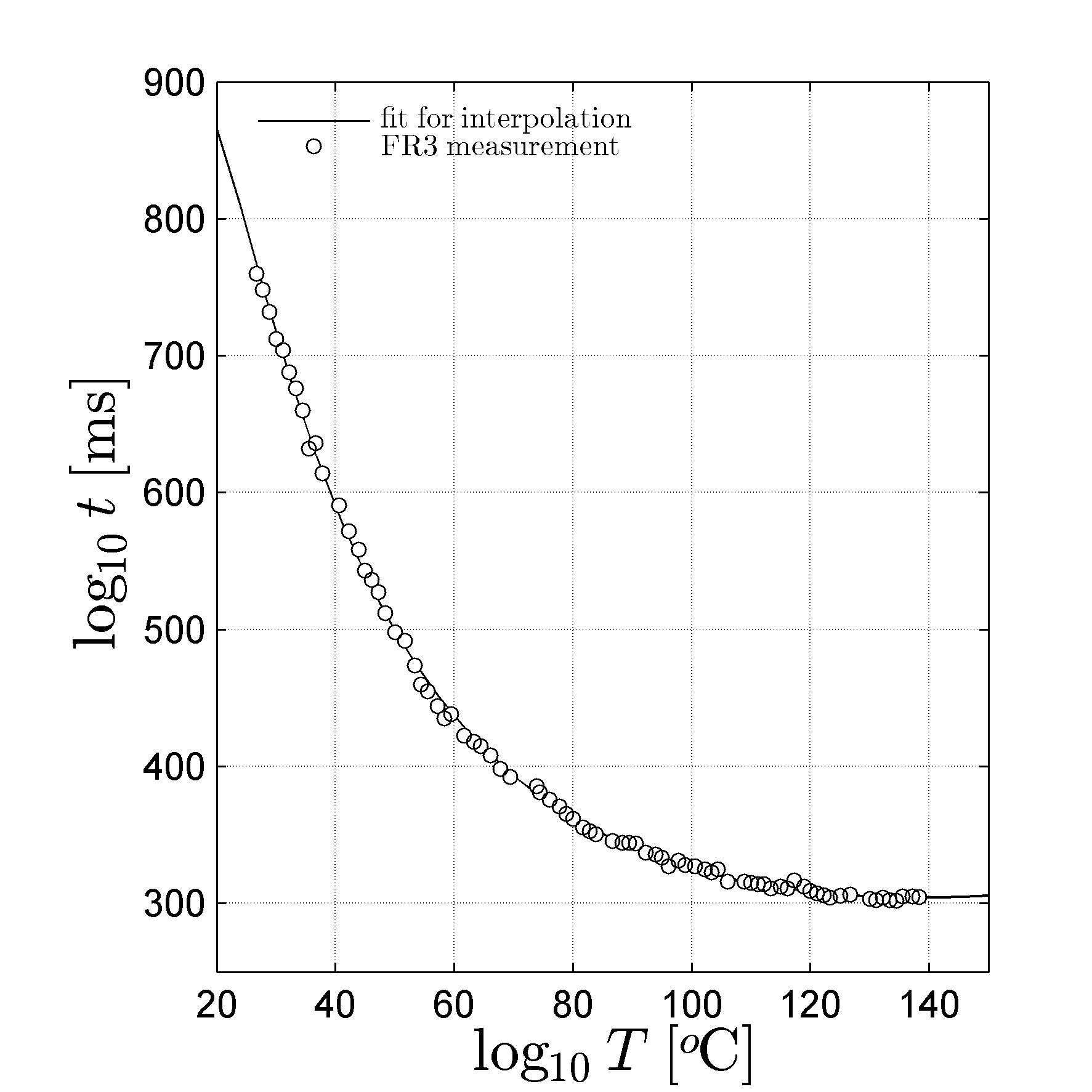}
  \caption{Time-to-fall measurement for FR3\texttrademark~ as a function of temperature measured in the autoclave under N$_2$ atmosphere. The solid  line is used for the transfer function.}
  \label{fig:tVStemp}
\end{figure}

\begin{figure}[htp]
  \centering
  \includegraphics[width=\linewidth]{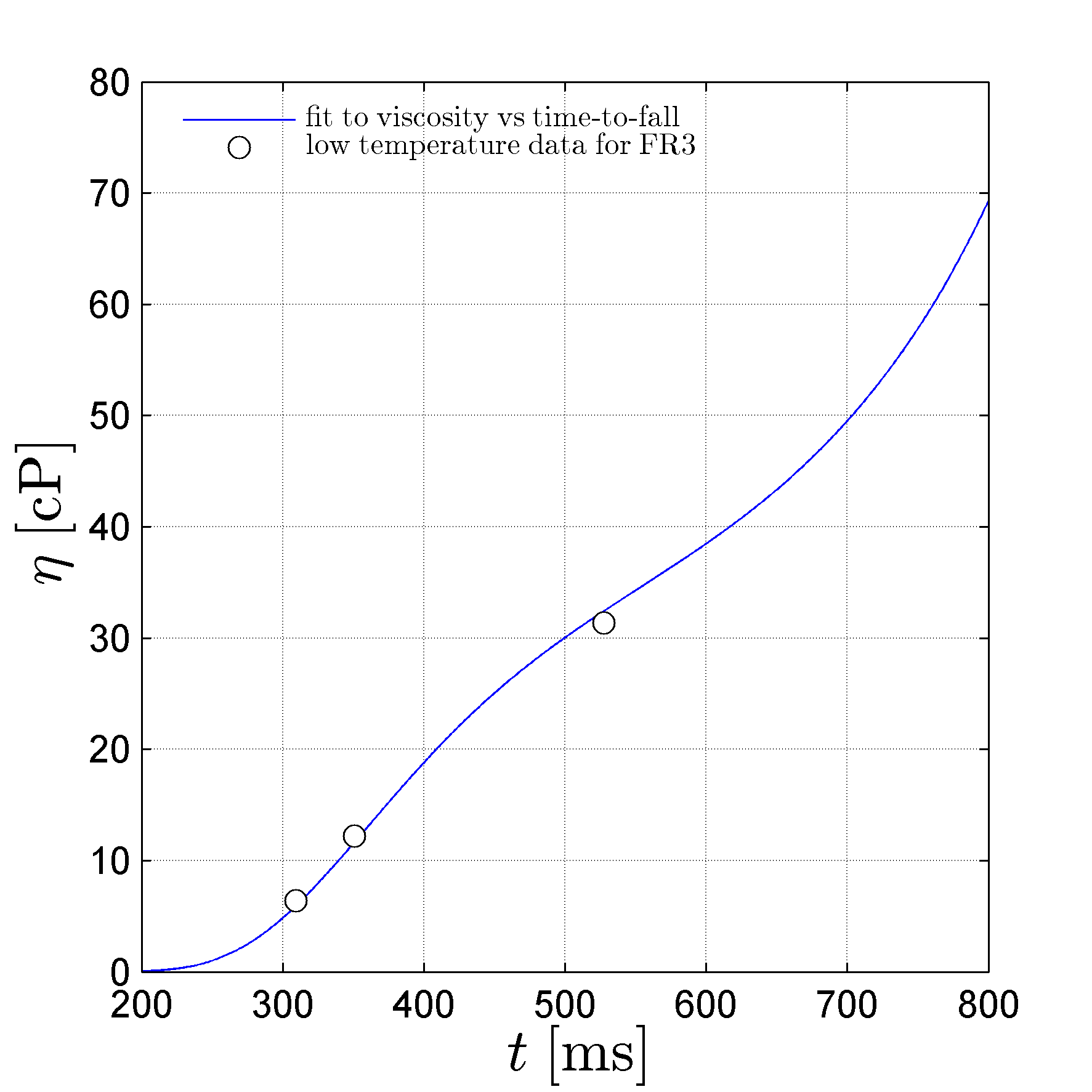}
  \caption{Time-to-fall to viscosity transfer function for FR3\texttrademark. The solid line is used to convert all our measurement to viscosity in cP. The data is shown with open symbols is for FR3\texttrademark~ from the low temperature conventional measurement.}
  \label{fig:etaVSt}
\end{figure}

\begin{figure}[htp]
  \centering
  \includegraphics[width=\linewidth]{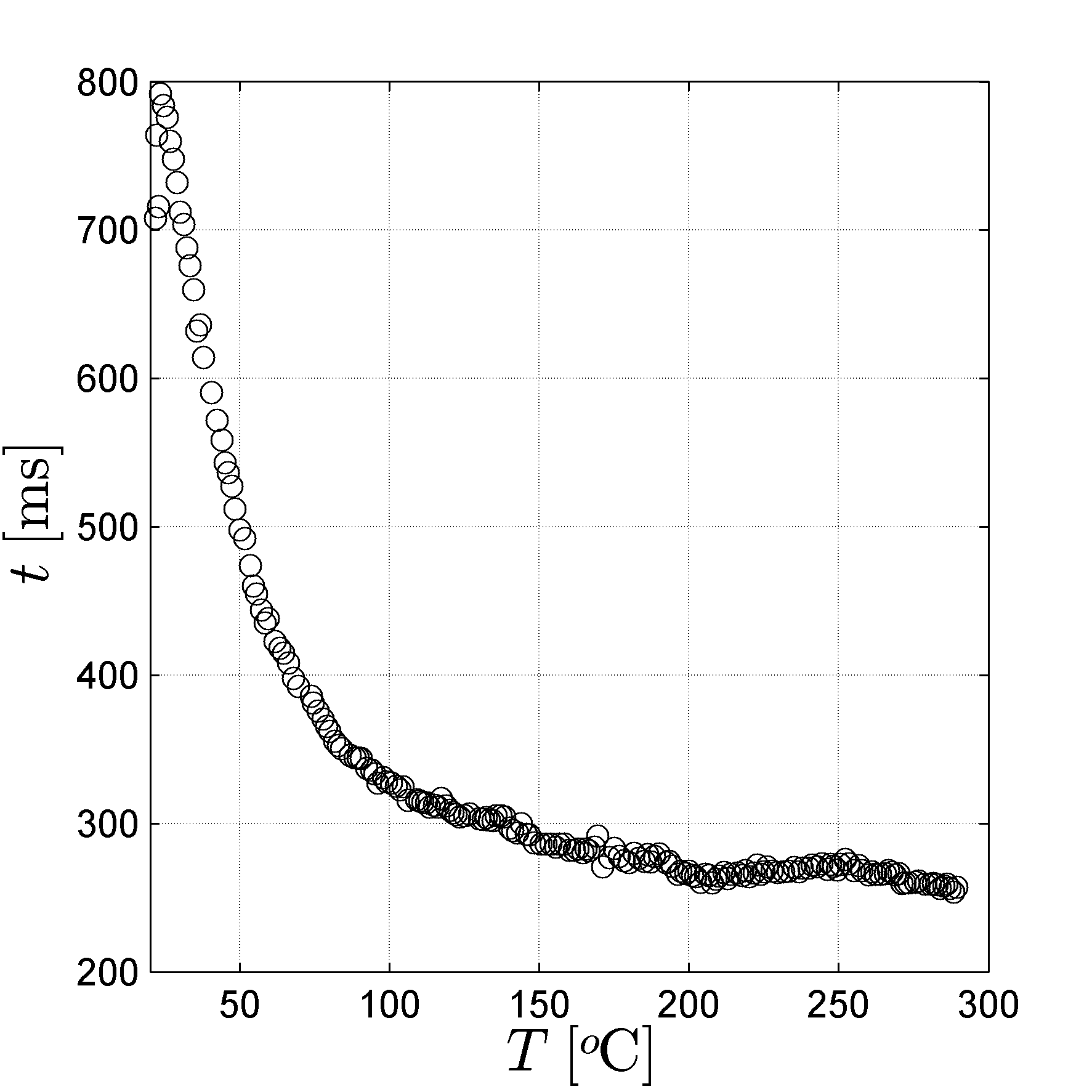}
  \caption{Time-to-fall as a function temperature for FR3\texttrademark.}
  \label{fig:tallVST}
\end{figure}

\begin{figure}[htp]
  \centering
  \includegraphics[width=\linewidth]{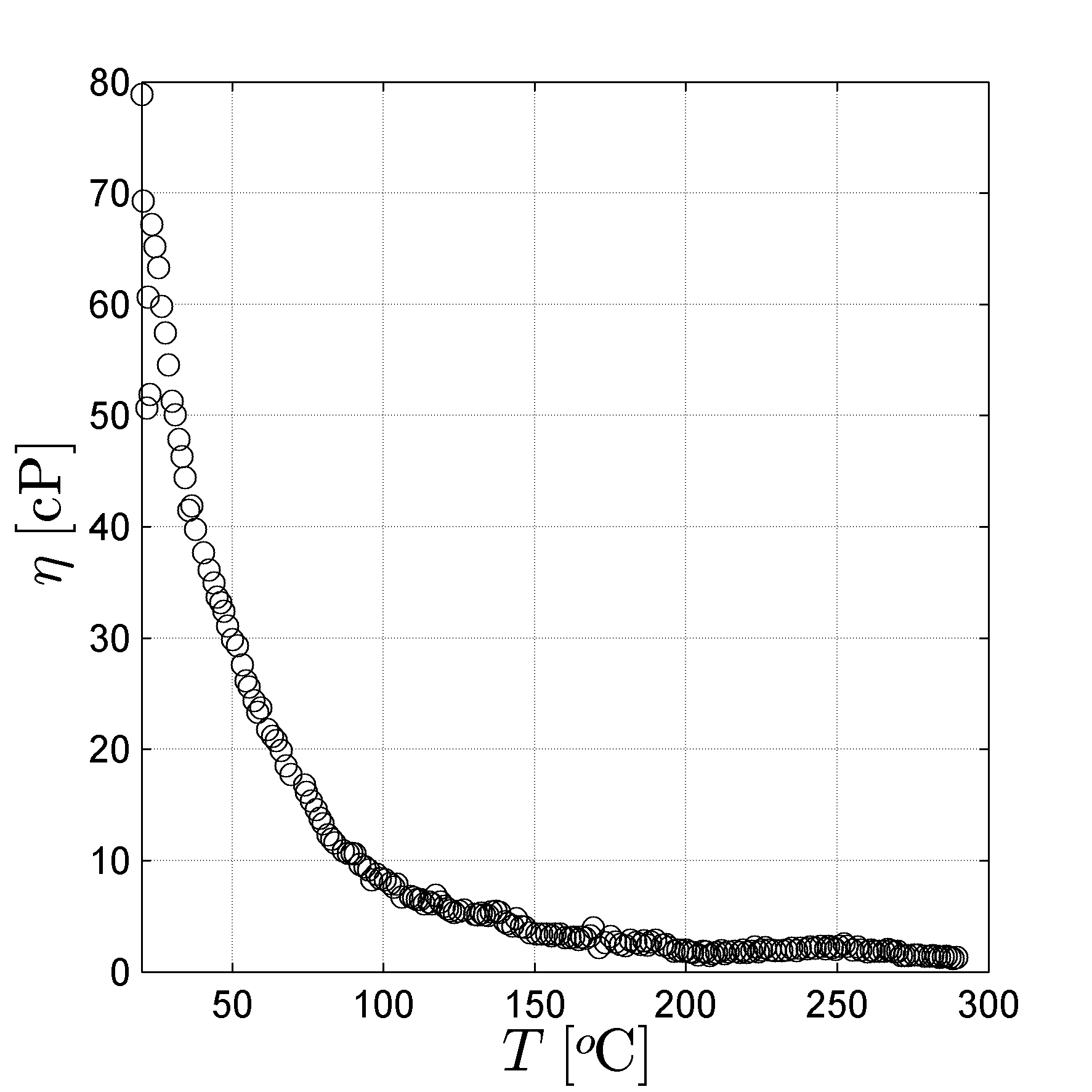}
  \caption{Viscosity estimates from time-to-fall measurements for FR3\texttrademark.}
  \label{fig:etaVST}
\end{figure}

\begin{figure}[htp]
  \centering
  \includegraphics[width=\linewidth]{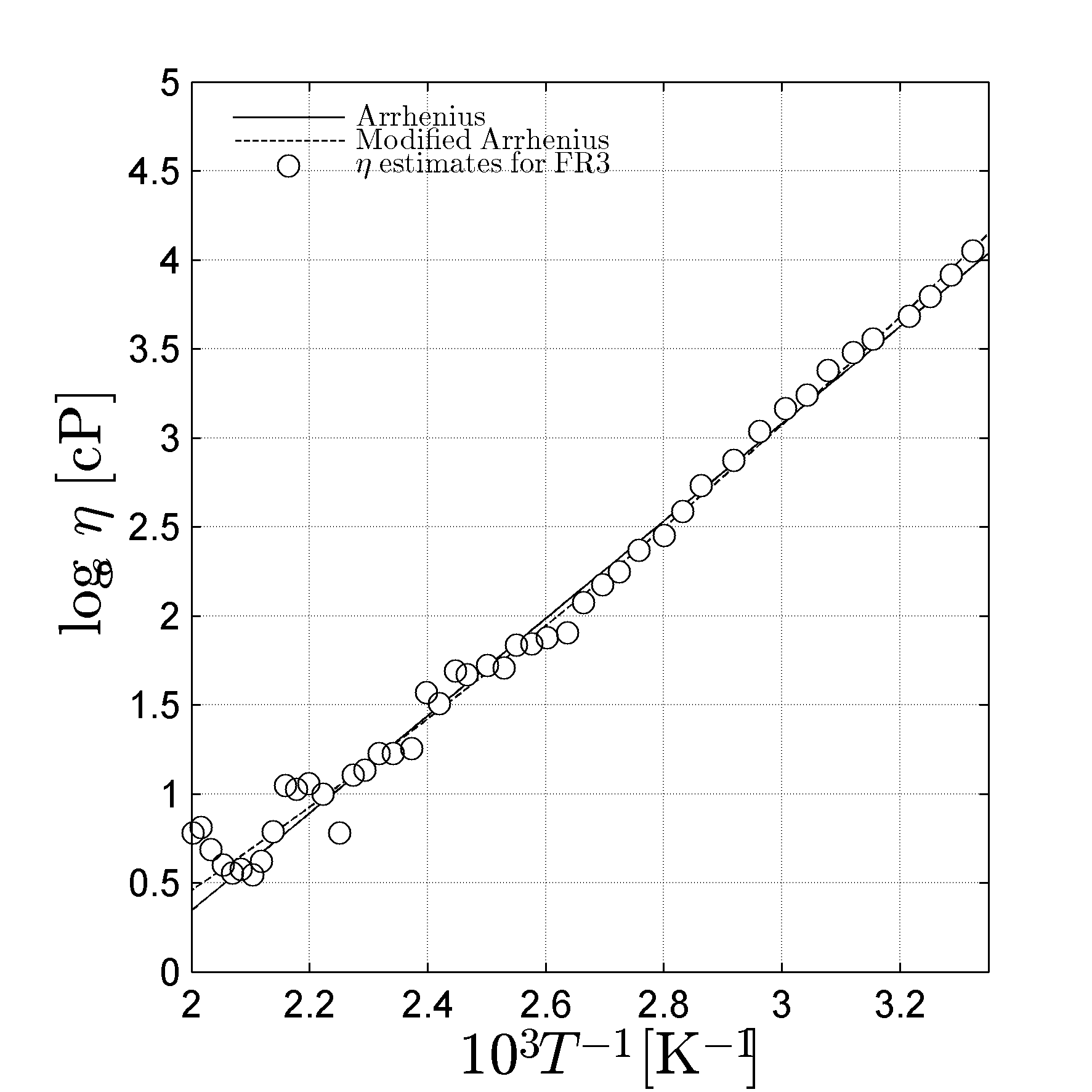}
  \caption{Estimated viscosity as a function of inverse absolute temperature for FR3\texttrademark. The solid and dashed lines represented the Arrhenius and modified Arrhenius fits to the estimates, respectively.}
  \label{fig:etaVSinvT}
\end{figure}

To test the measurement setup, and build the transfer function between measured time-to-fall and viscosity, an oil with existing viscosity data was selected\cite{GerstlerGE}. The oil is a commercially available transformer oil called Envirotemp\texttrademark~FR3\texttrademark~ Dielectric Fluid from Cargill Inc. USA. Its viscosity was previously\cite{GerstlerGE} measured up to 150$\degC$C using a conventional method. Its viscosity data are shown in Fig.~\ref{fig:etaVStemp}. The data are described with a log-log model as
\begin{eqnarray}
  \label{eq:RTconvent}
  \log_e\eta= -0.298\times(\log_eT)^2+    0.874\times\log_eT  +  4.50,
\end{eqnarray}
where $\eta$ is the viscosity and $T$ is the temperature in $\degC$C. Similarly the time-to-fall $t$ from our measurement is plotted in Fig.~\ref{fig:tVStemp}, where the measurements were performed in the autoclave under N$_2$ atmosphere. The curve fitted to our time-to-fall data $t$ is also a log-log model,
\begin{eqnarray}
  \label{eq:tvsT}
  \log_et= 0.215\times(\log_eT)^3-2.48\times(\log_eT)^2+8.82\times\log_eT-3.15
\end{eqnarray}
These two equations can be combined as a transfer function which is shown in Fig.~\ref{fig:etaVSt}. The data for low temperatures are also plotted in Fig.~\ref{fig:etaVSt}. Using the relationships in Eqs.~(\ref{eq:RTconvent}) and (\ref{eq:tvsT}),the conversion between time-to-fall $t$ in ms units and viscosity $\eta$ in cP is derived as below,
\begin{eqnarray}
  \label{eq:trans}
   \log_e\eta= 4.75\times(\log_et)^3	-90.1\times(\log_et)^3+572\times(\log_eT)	-1210.
\end{eqnarray}
The time-to-fall are then converted to viscosity. The whole data set for FR3\texttrademark~ are shown in Fig.~\ref{fig:tallVST}. The results for the estimated viscosities at different temperatures are shown in Fig.~\ref{fig:etaVST}. As expected the viscosity of the oil decreased with increasing temperature. The dependence of the oil viscosity to temperature is estimated with two expressions\cite{Fulcher},
\begin{eqnarray}
  \log_e \eta&=&a(k_bT)^{-1}+b  \label{eq:Arrhenius}\\
  \log_e \eta&=&AT^{-2}+BT^{-1}+C  \label{eq:Fulcher}
\end{eqnarray}
where $a$, $b$, $A$, $B$ and $C$ are fit parameters. The results are shown in Fig.~\ref{fig:etaVSinvT}. For the Arrhenius fit, Eq.~(\ref{eq:Arrhenius}), the parameters $a$ and $b$ are 2730 and -5.12, respectively. For the quadratic expression in Eq.~(\ref{eq:Fulcher}), $\{A,B,C\}$ are $\{342000,904,-2.72\}$. Eq.~(\ref{eq:Fulcher}) res presents the data better than Eq.~(\ref{eq:Arrhenius}), since it has a third unknown. 

\section*{Discussion and conclusions}
A device was developed to measure liquid viscosity properties as a function of temperature and pressure. The device functions as an insert for an autoclave where temperature and pressure were controlled. A straightforward fall time of a magnet was measured to estimate the viscosity of the liquid that the magnet was inserted without changing the initial charging pressure. The obtained time data was converted to viscosity by employing natural logarithmic dependencies due to unphysical nature of the negative time and viscosity. The relationship between time and viscosity were built using a known oil that was previously characterized in our labs at low temperatures. Estimated viscosity indicated an  Arrhenius behavior with an activation energy ($a\times ]k_B$) of 0.23$e$V, where $k_B$ is the Boltzmann coefficient, $k_B=8.613210^{-5}e$V. It was shown that the device was efficient to measure viscosity at elevated temperatures and pressures. Future studies will concentrate on other dielectric oils under high temperatures and pressures.  

\subsection*{Acknowledgment}
Drs Weijun Yin, John Krahn and Manoj Shah from GE GRC are thanked for the fruitful discussion and their directions in designing the measurement jig and setting-up the autoclave. This work was funded in part by the U.S. Department of Energy¡¯s Geothermal Technologies Program DE-EE0002752 titled `High-Temperature-High-Volume Lifting for Enhanced Geothermal Systems.'


\bibliographystyle{unsrtnat}
\end{document}